\documentclass[12pt]{iopart}

\usepackage{graphicx}

\usepackage{iopams}

\begin{document}

\title[The electron internal clock]
{Measuring the internal clock of the electron}

\author{Mart\'{\i}n Rivas}
\address{Theoretical Physics Department, The University of the Basque Country,\\ 
Apdo.~644, 48080 Bilbao, Spain}
\ead{martin.rivas@ehu.es}

\begin{abstract}
The existence of an internal frequency associated to any elementary 
particle conjectured by de Broglie is compared
with a classical description of the electron, where this internal structure corresponds to the
motion of the centre of charge around the centre of mass of the particle. This internal motion
has a frequency twice de Broglie's frequency, which corresponds to the frequency found by Dirac
when analysing the electron structure. To get evidence of this internal electron clock a 
kind of experiment as the one performed by Gouan\'ere et al. \cite{Gouanere} will show a discrete
set of momenta at which a resonant scattering effect, appears. 
The resonant momenta of the electron beam are given by $p_k=161.748/k$ MeV$/c$, $k=1,2,3,\ldots$,
where only, the corresponding to $k=2$, was within the range of Gouan\'ere et al. experiment.
The extension of the experiment to other values of $p_k$, would show the existence of this phenomenon.
\end{abstract} 


\maketitle

\section{Introduction}\label{sec:intro}
The main conjecture in de Broglie's thesis is {\ldots\it the existence of a periodic phenomenon
of an unknown nature, associated to every portion of isolated energy, and related to 
its proper mass by Planck-Einstein's equation}  \cite{Broglie}, so that a particle of mass $m$ at rest should have an internal
frequency $\nu_0=mc^2/h$ and when the particle moves at a velocity $v$ this frequency will be $\nu=mc^2/h\gamma$,
where $\gamma=(1-v^2/c^2)^{-1/2}$ is the relativistic time dilation factor. Properly speaking, a moving electron
has a greater energy $\gamma mc^2$ and therefore a greater frequency $\gamma mc^2/h$ according to the above statement, 
but if this internal periodic motion has a physical reality will have the frequency proposed by de Broglie $\nu=mc^2/h\gamma$
when measured by an inertial observer who sees the electron moving at the speed $v$, and according to 
the time dilation of special relativity.

\begin{figure}
\begin{center}\includegraphics[width=7cm]{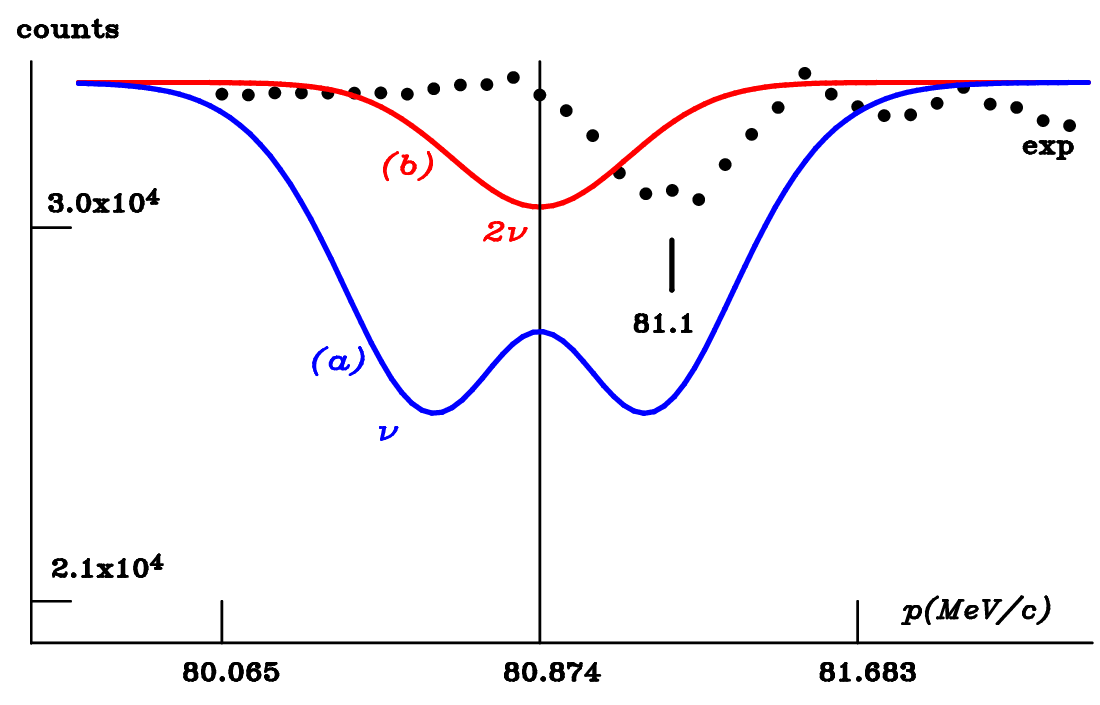}\end{center}
\caption{\label{fig:1}{Transcription of figure 4 of 2008 reference \cite{Gouanere}, which shows
the experimental outcome of the detected number of electrons versus the linear momentum $p$ of the electron beam
in MeV$/c$ (dotted line).
Curve (a) represents their Monte Carlo calculation for
de Broglie's frequency $\nu$.
Curve (b) represents their Monte Carlo calculation by considering that the electron internal frequency is twice
de Broglie's frequency $2\nu$. It matches with the experimental result except for a shift from $80.874$ MeV$/c$ to $81.1$ MeV$/c$}} 
\end{figure}

Using this conjecture, Gouan\`ere et al. \cite{Gouanere} tried to get evidence of this electron internal clock
frequency by analysing the interaction of a beam of electrons with the atoms of a $1\mu$m thick  
silicon crystal  aligned along the
$<110>$ direction. If the momentum of the electron beam is the appropiate, they will interact in a kind of 
transversal resonant scattering, whenever the distance travelled by the electron during a 
period $L= v/\nu$ will be related to the
separation $d=3.84${\AA} between the atoms of the crystal. 
When
\[
L=\frac{v}{\nu}=\frac{h\gamma v}{mc^2}=\frac{hp}{m^2c^2}=d
\]
this will produce a lateral force such that the 
outgoing beam in the forward direction will show a decrease in the number of outgoing electrons for that
resonant momentum $p$. This resonant momentum is $p=80.874$ MeV$/c$ while their experiment
shows a value of $p_{exp}=81.1$ MeV$/c$, a 0.28\% higher than the predicted theoretical value (see Figure~\ref{fig:1}).

Their Monte Carlo calculation, using parabolic potentials to describe the atomic structures, 
shows a bad fitting with the  
experimental result for this internal frequency (curve (a) of Figure~\ref{fig:1}), while
a clear matching (curve (b) of Figure~\ref{fig:1}), except for the above 0.28\% shift, 
between the expected and the experimental outcome 
for a momentum range between $80.0$ MeV$/c$ and $82.0$ MeV$/c$ and when the
electron frequency is taken twice de Broglie's frequency $2\nu$ or $v/2\nu=d$. This resonant momentum
is $161.748$ MeV$/c$, it is outside of the experiment range, 
and the observed peak corresponds to the second harmonic. 
This frequency corresponds to the
zitterbewegung frequency or frequency found by Dirac of the internal motion of the electron.

We are going to produce this analysis based upon a classical electron model found from a very general formalism
for describing spinning particles \cite{Rivasbook}. Recently \cite{carga}, we have given different arguments justifying that
the centre of charge and centre of mass of a spinning particle are different points, and, for the electron at rest,
the centre of charge is moving in circles, at the speed of light, around the centre of mass with a frequency
$\nu_0=2mc^2/h$, twice as much as de Broglie's frequency and coherently matched with Dirac's internal
frequency. This internal frequency changes as $\nu=\nu_0/\gamma$ when the electron moves, because of the special 
relativity time dilation. 
It is this relative motion of the centre of charge which has a clear interpretation as the 
Schroedinger's zitterbewegung, and therefore a clear internal periodic phenomenon associated to a spinning particle. 


\section{The classical electron model}
In the kinematical formalism \cite{Rivasbook}, an elementary particle is, by definition, 
a mechanical system which in addition to being indivisible, as a consequence 
of the atomic hypothesis  \cite{atomic}, 
it can never be deformed so that all allowed states
are only kinematical modifications of any one of them. This means that when the state
of an elementary particle changes there exists another inertial observer
who measures the particle in the same state as before. This means that in a variational 
approach of classical physics, 
the initial $x_1$ and final $x_2$ states of the evolution of an elementary particle 
must be related by a transformation of the
kinematical group $x_2=gx_1$. Therefore, the boundary variables
of the variational approach, necessarily span a homogeneous space of the kinematical 
group of space-time symmetries. When quantizing all classical systems 
characterized by such homogeneous spaces, their Hilbert space of pure states carries a projective
unitary irreducible representation of the kinematical group \cite{RivasQ}. It thus satisfies Wigner's
definition of a quantum elementary particle.

In this way, the parameters of the kinematical group become the classical variables we use,
as the boundary values of the variational formalism for describing an elementary particle. 
In the relativistic and non-relativistic approach, these
variables are reduced to the ten variables $t,{\bi r},{\bi u}$ and $\balpha$, interpreted respectively 
as the time, position of the charge, velocity of the charge and orientation of the particle.
In the relativistic case we have three
disjoint, maximal homogeneous spaces of the Poincar\'e group 
spanned by these variables with the constraint either $u<c$, $u=c$
or $u>c$. It is the manifold with $u=c$, as suggested by the kinematical analysis of reference \cite{carga}, 
which leads to Dirac's equation when quantizing the system. This kinematical analysis states the possibility
that the centre of mass and centre of charge of a spinning particle could be different points. In that case,
the centre of charge has necessarily a velocity of constant absolute value, 
unreachable by any inertial observer, so that only the relativistic approach
is selected and the charge moves at the speed $c$.

The Lagrangian depends on these ten kinematical variables and also on the 
acceleration of the point ${\bi r}$ and on the angular velocity. 
For a Dirac particle, the charge located at point ${\bi r}$
is moving at the speed of light $u=c$. 
The classical expression linear in the energy $H$ and in the linear momentum ${\bi P}$, 
which gives rise to Dirac's equation is 
\[
H={\bi P}\cdot{\bi u}+\frac{1}{c^2}{\bi S}\cdot\left(\frac{d{\bi u}}{dt}\times{\bi u}\right),
\]
where the energy $H$ is expressed as the sum of two terms, ${\bi P}\cdot{\bi u}$, or translational energy
and the other, which depends on the spin of the system, or rotational energy. 
The spin comes from the dependence of the Lagrangian $L$
of both, the acceleration $\dot{\bi u}$, and the angular velocity  ${\bomega}$, and if we define
\[
{\bi U}=\frac{\partial L}{\partial \dot{\bi u}},\quad {\bi W}=\frac{\partial L}{\partial\bomega},
\] 
it takes the form
\[
{\bi S}={\bi u}\times{\bi U}+{\bi W}={\bi Z}+{\bi W}.
\]
The first part ${\bi Z}={\bi u}\times{\bi U}$, or {\it zitterbewegung} part, 
is related to the separation between the centre of charge from the centre of mass and 
takes into account this relative orbital motion. It quantizes with integer values. 
The second part ${\bi W}$
is the rotational part of the body frame and quantizes with both integer and half-integer values.
The total angular momentum with respect to the origin of observer's frame is
\[
{\bi J}={\bi r}\times{\bi P}+{\bi S},
\]
so that the spin ${\bi S}$ is the angular momentum of the system with respect 
to the centre of charge ${\bi r}$,
and not with respect to the centre of mass ${\bi q}$. By this reason, it
is not a conserved quantity for a free particle,
but satisfies the dynamical equation
 \begin{equation}
\frac{d{\bi S}}{dt}={\bi P}\times{\bi u}.
 \label{eq:dynspin}
 \end{equation}
This is the same dynamical equation satisfied by Dirac's spin operator in the quantum case.

\begin{figure}
\begin{center}\includegraphics[width=7cm]{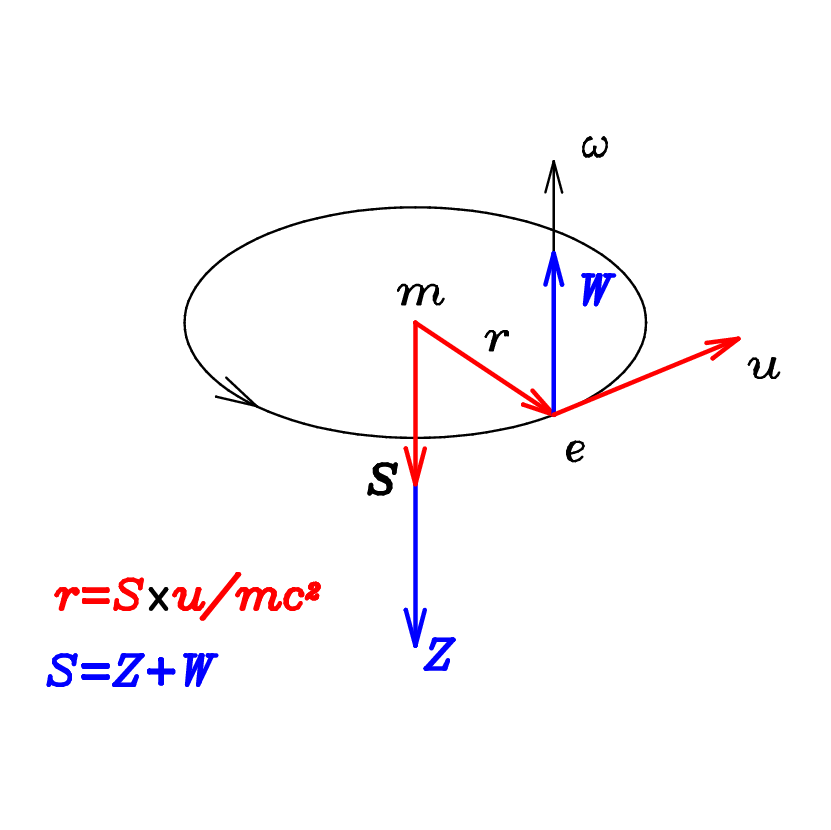}\end{center}
\caption{\label{fig:3}{Motion of the charge of the electron at the speed of light in the centre of mass frame. The magnetic moment
of the particle is produced by the motion of the charge. The frequency of this internal motion in this frame is 
$\nu_0=2mc^2/h$. The total spin
${\bi S}$ is half the value of the zitterbewegung part ${\bi Z}$ when quantizing the sytem, so that when expressing the magnetic
moment in terms of the total spin we get a $g=2$ gyromagnetic ratio \cite{g2}. 
The body frame attached to the point ${\bi r}$
rotates with angular velocity $\omega=2\pi\nu_0$, has not been depicted.}} 
\end{figure}

When expressed Dirac's spin ${\bi S}$ and the centre of mass position ${\bi q}$ 
in terms of the velocity and acceleration of the charge they take, respectively, the form
\[
{\bi S}=\left(\frac{H-{\bi u}\cdot{\bi P}}{\left({d{\bi u}}/{dt}\right)^2}\right)\,\frac{d{\bi u}}{dt}\times{\bi u},\quad 
{\bi q}={\bi r}+\frac{c^2}{H}\left(\frac{H-{\bi u}\cdot{\bi P}}{(d{\bi u}/dt)^2}\right)\frac{d{\bi u}}{dt}
\]
Dirac's spin is always orthogonal to the osculator plane of the trajectory 
of the charge ${\bi r}$, 
in the direction opposite to the binormal for a positive energy particle, and in the 
opposite direction for the antiparticle. This implies a difference in chirality between matter and 
antimatter. Point ${\bi q}$ is always different from point ${\bi r}$, because the observable
$H-{\bi u}\cdot{\bi P}\neq0$ thus justifying the conjecture that for a spinning particle the centre
of mass and centre of charge are different points.
The acceleration of the charge is pointing from ${\bi r}$ to the centre of mass ${\bi q}$, 
as it corresponds to a helix. 
It is shown that the dynamical equation of point ${\bi r}$ for the free particle and 
in the centre of mass frame is given by
 \begin{equation}
{\bi r}=\frac{1}{mc^2}{\bi S}\times{\bi u},
 \label{eq:dyq}
 \end{equation}
where the spin vector ${\bi S}$ is constant in this frame, as depicted in Fig.~\ref{fig:3}.
The radius of the zitterbewegung motion is $R=S/mc$, and the angular velocity $\omega=mc^2/S$.
When quantizing the system the classical parameter $S=\hbar/2$.
The frequency of this internal motion and in the centre of mass frame 
is $\nu_0=2mc^2/h$, i.e., twice de Broglie's frequency, and the radius is $R=\hbar/2mc$, is half Compton's wavelength.

\begin{figure}
\begin{center}\includegraphics[width=7cm]{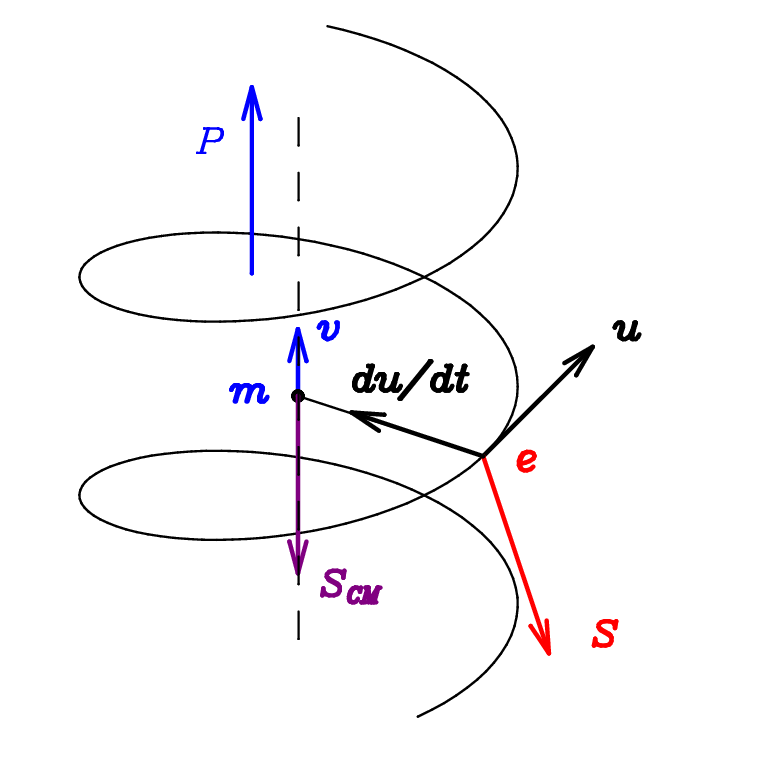}\end{center}
\caption{\label{fig:4}{Precession of Dirac's spin ${\bi S}$ along the linear momentum ${\bi P}$ for a moving electron. 
The tranversal motion of the charge takes a time $\gamma(v)$ longer than when the centre of mass is 
at rest, to complete a turn and, therefore, the frequency of the internal motion of the charge is 
$\nu=\nu_0/\gamma$.
The spin with
respect to the centre of mass ${\bi S}_{CM}$, is a constant of the motion for a free particle.}} 
\end{figure}

When seen from an arbitrary observer (see Figure~\ref{fig:4}), the motion of the charge for a longitudinaly
polarized electron is a helix, 
so that according to (\ref{eq:dynspin}) Dirac's spin precess around the direction of the 
conserved linear momentum ${\bi P}$. The spin with respect to the centre of mass is defined as
\[
{\bi S}_{CM}={\bi S}+({\bi r}-{\bi q})\times{\bi P}.
\]
It is a conserved quantity for a free particle. 
The centre of mass velocity is ${\bi v}=d{\bi q}/dt$, 
and the linear momentum is written as usual
as ${\bi P}=\gamma(v)m{\bi v}$. This means that the transversal motion of the charge
is at the velocity $\sqrt{c^2-v^2}$. The centre of charge of a moving 
electron takes a time $\gamma(v)$ times longer
than for an electron at rest to complete a turn, which is a clear consequence 
of the time dilation.
The faster the centre of mass of the electron moves the slower is the rotation frequency 
of the centre of charge around the centre of mass. The internal clock of a fast electron is running
slower and we are going to take advantage of this fact to get the resonant interaction with a lattice of 
atoms. The simple idea is that if it exists this temporal periodic motion of the centre of charge around the centre
of mass, when the particle moves with constant velocity we also get a spatial periodicity. If this spatial periodicity
is confronted with another spatial periodic structure, like a crystal, and produce the interaction with every atom of the lattice,
there will be a resonant scattering for some selected momenta.

\section{Polarized beams}
\begin{figure}
\begin{center}\includegraphics[width=7cm]{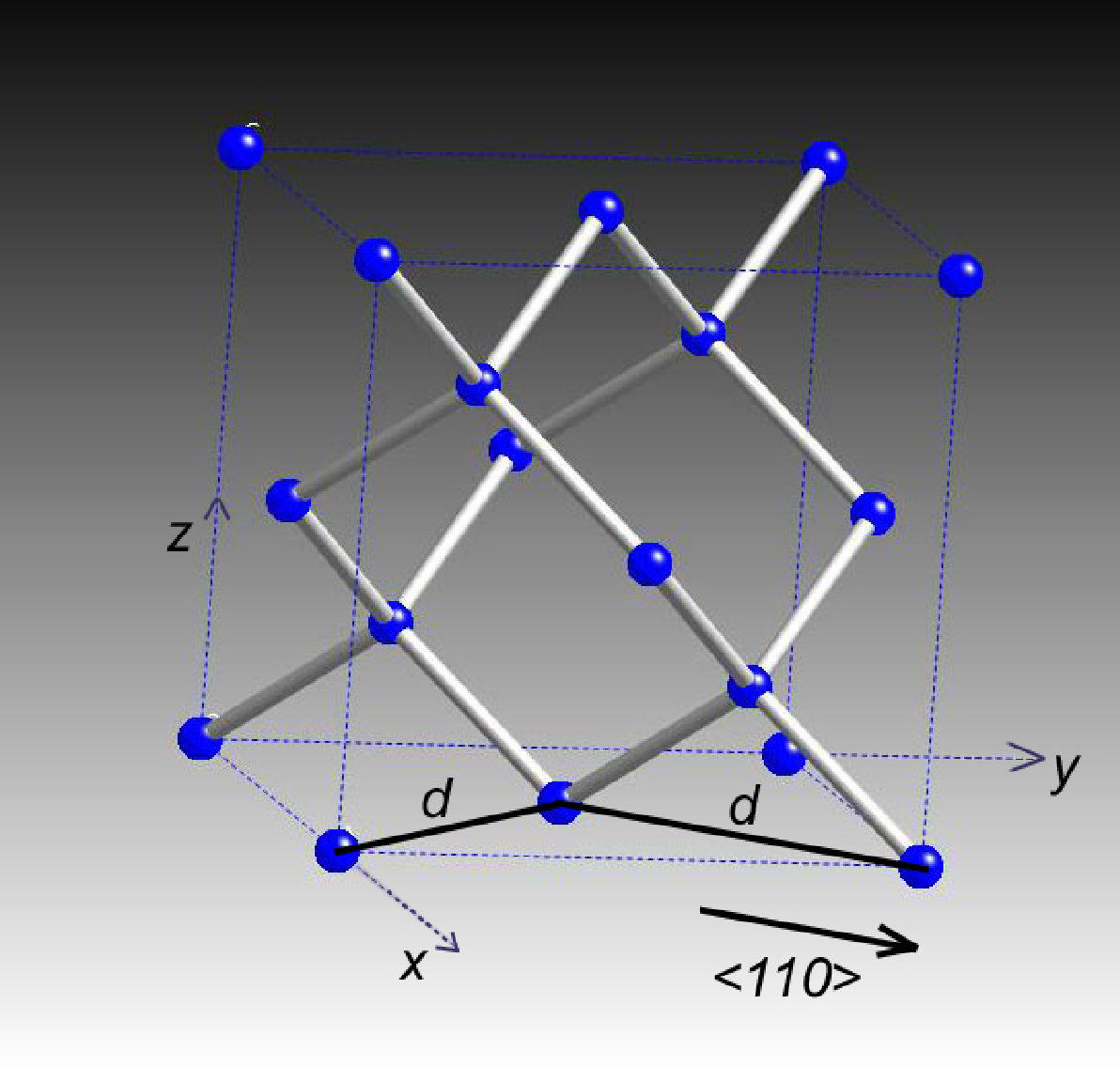}\end{center}
\caption{\label{si}{Silicon crystal unit cell, showing the direction $<110>$ on the $XY$ plane, and 
the separation $d=3.84${\AA} between neighbouring atoms on that plane.}}
\end{figure}

\begin{figure}
\begin{center}\includegraphics[width=7cm]{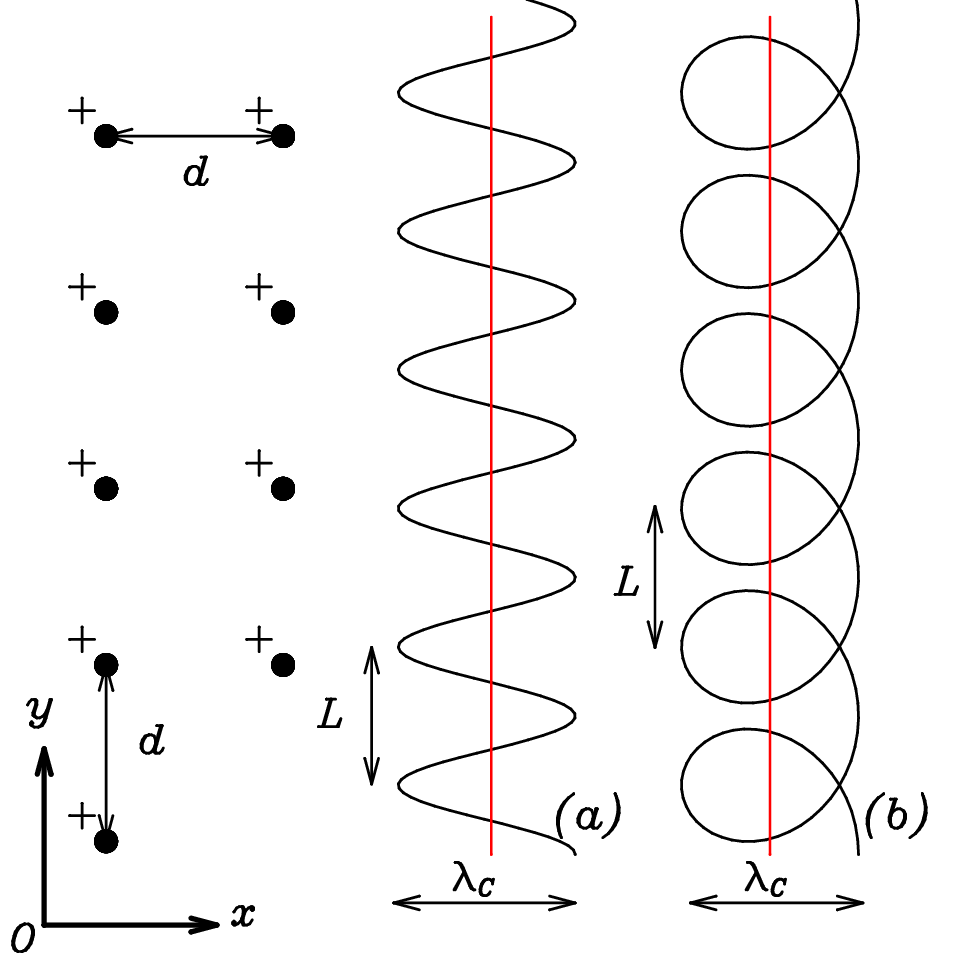}\end{center}
\caption{\label{fig:2}{Part (a) describes a free motion of an electron polarized longitudinally along the $OY$ axis, 
while part (b) describes the free motion of an electron polarized transversally in the direction $OZ$, 
perpendicular to the picture and the motion of the centre of charge is contained on the $XY$ plane. 
The straight line represents the centre of mass motion and the curly line is the trajectory described by the centre of charge. 
The range $\lambda_C=3.86\times10^{-13}$m of the transversal motion of the centre of charge is Compton's wave length,
and $d=3.84\times10^{-10}$m represents the separation between atoms of the Si crystal on the plane
corresponding to the $<110>$ direction of the electron beam. 
Both lengths are clearly not represented at the same scale.
$L$ is the distance travelled by the electron during a period. There will be resonant scattering
whenever $L=nd$ or $d=kL$, for $n,k=1,2,3,\ldots$}} 
\end{figure}

In figure \ref{si} we show the unit cell of a silicon crystal and the separation $d$ between
neighbouring atoms on the $XOY$ plane along which the electron beam is sent in the Gouan\'ere et al. experiment \cite{Gouanere}.
In the figure \ref{fig:2} we show two possible free motions of a polarized electron which show a clear
spatial periodicity. In part (a) we describe
the motion of the centre of mass and centre of charge of an electron polarized along the direction of motion $OY$. 
In part (b) the electron is polarized perpendicularly to the page along the $OZ$ axis. 
In both cases the centre of mass motion is in the
direction $<110>$ of the crystalographic plane, where the dots represent the location of the Si nuclei on that plane.
For electrons polarized along the $OX$ direction, where the centre of charge motion is contained
in the plane $YOZ$, the description is equivalent to the last one, where the atoms will be arranged
on the $YOZ$ plane.
The frequency of the internal motion of a moving electron is $\nu=2mc^2/h\gamma$, so that the distance
travelled by the centre of mass during a period is $L=h\gamma v/2mc^2=hp/2m^2c^2$. There will be a resonant
scattering whenever $L=nd$ or when $d=kL$ for any natural number $n,k=1,2,3,\ldots$. In the first case, the particle
moves faster and there will be a resonant interaction every $n-$th atom, while in the second case
the momentum will be smaller, the particle will interact with each atom but 
every $k$ periods of the electron.
The previsible values of these momenta are
\[
p_n=n\frac{2m^2c^2d}{h},\quad p_k=\frac{1}{k}\frac{2m^2c^2d}{h},\quad n,k=1,2,3,\ldots
\]
\[
p_n=161.748,\quad 323.496,\quad 485.244,\quad 646.992,\ldots {\rm MeV/}c
\]
\[
p_k=161.748,\quad 80.874,\quad 53.916,\quad 40.437,\quad 32.3496,\quad
26.958,\ldots {\rm MeV/}c
\]
We can produce a numerical computation of the interaction of the incoming electron beam with the electrostatic nuclei potential 
or with the outer shells of the atoms. In any case, if there is a matching of $L$ with $d$, as suggested above, 
the amount of transversal momentum transfer $\Delta p_x$ to the electron from each atom, 
either to the right or to the left, 
will be of the same sign with each atom, so that the total transversal momentum transfer will be $N\Delta p_x$,
after $N$ atoms.
In the Gouan\'ere et al. experiment \cite{Gouanere}
the range of the electron beam was prepared between 54 to 110 MeV$/c$, so that
only the resonant frequency for $k=2$,  80.874 MeV$/c$ was available within that range, although they 
mention that {\it an observation could also be possible at harmonic frequencies, for example
at corresponding momenta of {\rm 161.748 MeV}$/c$ and {\rm 40.437 MeV}$/c$, etc., probably with reduced intensity.}. 

\begin{figure}
\begin{center}\includegraphics[width=10cm]{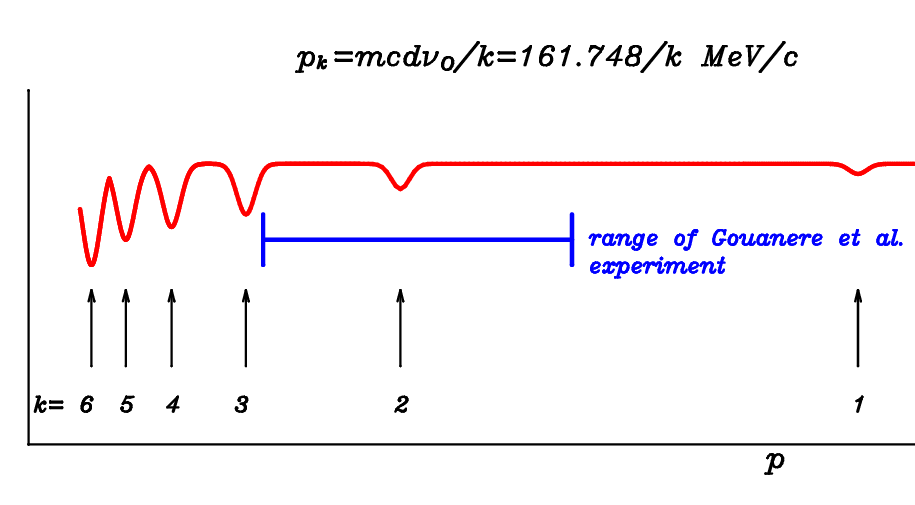}\end{center}
\caption{\label{Fig:Dirac}{Different resonant peaks of the interaction of the electron beam with the silicon 
latice, if assumed that the internal electron frequency is twice De Broglie's frequency $\nu_0=2mc^2/h$. Only the
peaks corresponding to $p_k$, $k=1,\ldots,6$, are depicted.}}
\end{figure}

\begin{figure}
\begin{center}\includegraphics[width=10cm]{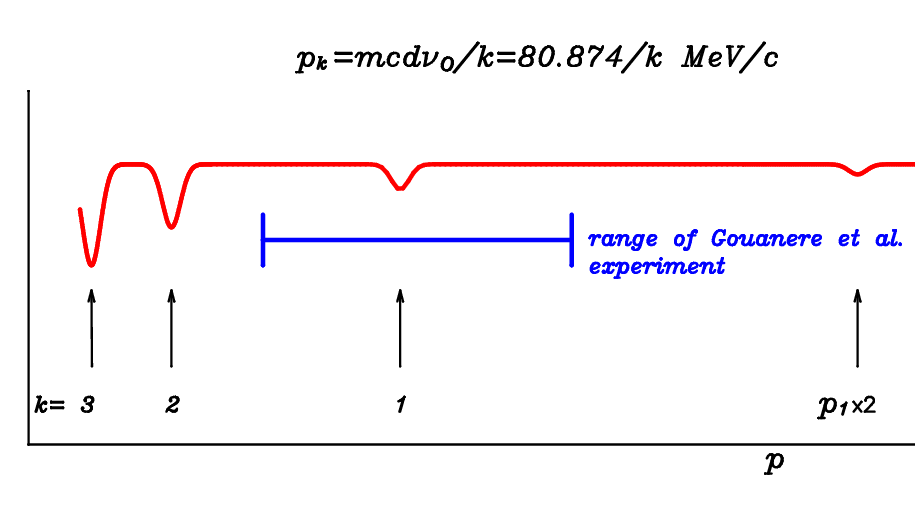}\end{center}
\caption{\label{Fig:Broglie}{Different resonant peaks of the interaction of the electron beam with the silicon 
latice, if assumed that the internal electron frequency is De Broglie's frequency $\nu_0=mc^2/h$. Some of the previsible peaks
of the previous figure do not appear in this ansatz.}}
\end{figure}

Our conjecture is that the momenta of small value $p_k$ are more favorable to produce
a larger transversal scattering because the velocity of the beam is smaller, the particle remains
in the crystal a longer time and the interaction is produced with every atom, 
and, therefore, the ratio of the transversal momentum transfer to the longitudinal 
momentum of the beam is larger, so that a greater number of electrons
will be withdrawn from the forward direction, while for the larger 
momenta $p_n$ the ratio of the transversal perturbation to the longitudinal velocity is smaller 
and only a smaller number of atoms participate in the resonant scattering. This means that smaller
the momenta the larger the depth of the counts of the resonant peak.

In the figures (\ref{Fig:Dirac}) and (\ref{Fig:Broglie}) we show the different resonant peaks that will appear
if the range of the experiment is enlarged, provided the internal frequency of the electron at rest is Dirac's frequency
$\nu_0=2mc^2/h$ or just De Broglie's frequency $\nu_0=mc^2/h$, repectively. In the second case, several peaks
of the previous figure will not appear. The depths of the peaks are only a rough estimate, since 
smaller the incoming momentum $p_y$, the greater the deflection angle of the beam. Therefore, the enlargement
of the range of the experiment to the region of low momenta will show the existence or not of the previsible
peaks, and the clarification of whether the internal frequency is Dirac's or De Broglie. An accurate measurement
of the momenta of the resonant peaks will produce an accurate measurement of the electron internal clock frequency $\nu_0$.

For very small $p_k$, and depending on the experimental resolution, it will be difficult to discriminate
among the corresponding resonance peaks. For example, for $k=99,100,$ and 101, the corresponding incoming
resonant momenta are
\[
p_{99}=1.63382 {\rm\; MeV}/c,\quad p_{100}=1.61748 {\rm\; MeV}/c,\quad p_{101}=1.60147 {\rm\; MeV}/c
\]
which differ from each other of around $0.01$ MeV$/c$ which seems to be smaller than the resolution
of the experiment.

\ack{
I thank my colleague J M Aguirregabiria for the use of his
Dynamics Solver program \cite{JMA} with which the numerical computations of the electron trajectories
have been done. I also thank Gotzon Madariaga for his help with the silicon structure.
This work was supported by The University of the Basque Country
(Research Grant~9/UPV00172.310-14456/2002).}


\end{document}